\documentclass[12pt]{iopart}

\expandafter\let\csname equation*\endcsname\relax
\expandafter\let\csname endequation*\endcsname\relax
\usepackage{amsmath}
\usepackage{upgreek}
\usepackage{siunitx} 
\usepackage{amssymb}
\usepackage{multirow}
\usepackage{graphicx}
\usepackage{cite}
\usepackage{graphicx}
\usepackage{float}
\usepackage{booktabs}
\usepackage{caption}

\begin{document}

\title[Manuscript]{Evaluation of the effect of edge cracks on critical current degradation in REBCO tapes under tensile stress}

\author{Zhirong Yang$^{1}$, Peng Song$^{1}$, Mingzhi Guan$^{2}$, Feng Feng$^{3}$, Timing Qu$^{1,*}$}

\address{1. State Key Laboratory of Tribology, Department of Mechanical Engineering, Tsinghua University, Beijing 100084, China}
\address{2. Institute of Modern Physics, Chinese Academy of Sciences, Lanzhou 730000, China}
\address{3. Lab of Intelligent Manufacturing and Precision Machining, Tsinghua Shenzhen International Graduate School, Tsinghua University, Shenzhen 518055, China}
\ead{tmqu@mail.tsinghua.edu.cn}

\vspace{10pt}
\begin{indented}
\item[]Oct 2021
\end{indented}

\begin{abstract}

The slitting process used for fabrication of $\rm REBa_2Cu_3O_x$ (REBCO, RE=Rare earth) tapes of required width will greatly improve production efficiency and reduce production costs. However, edge cracks induced by the slitting process of wide REBCO tapes may cause the premature degradation under a extremely high hoop (tensile) stress in high-field magnets. It is necessary to evaluate the edge cracks of REBCO tapes on the critical current ($I_{\rm c}$) degradation. This work aims to evaluate the effect of edge cracks on the $I_{\rm c}$ performance under tensile stress. $I_{\rm c}$ degradation under artificial cracks was measured to validate the applicability of linear elastic fracture mechanics for the REBCO film. Linear elastic fracture mechanics was used to get the mixed stress intensity factor of multiple edge oblique cracks. A model considering edge crack properties angle $\beta$, spacing $d$, and length $a$ is constructed to evaluate the critical load and critical cracks properties. When the stress intensity factor at the crack tip is less than $K_{\rm Ic}=2.3$ $\mathrm{MPa\sqrt{m}}$, edge cracks remain stable and do not propagate. Two kinds of REBCO tapes fabricated by different companies are evaluated, and cracks of these tapes will not cause premature degradation. This model could be used to evaluate the operation range of REBCO tapes and improve the manufacturing process.

\noindent{\it Keywords\ }: REBCO, degradation, stress intensity factor, multiple edge oblique cracks, fracture toughness

\end{abstract}

\section{Introduction}

The $\rm REBa_2Cu_3O_x$ (REBCO, RE=Rare earth) high-temperature superconductors with higher critical current density is an ideal material for high-field magnets \cite{Senatore2014,Maeda2014,Fietz2013,Bruzzone2018,van2013,Xiong2013,Wang2017,2018feng}. A extremely high hoop stress (also tensile stress) caused by Lorentz force in high-field magnets easily triggers the failure of REBCO tapes. Slitting the wide REBCO tape to the width required by the application can greatly reduce the cost of conductor manufacturing. However, edge cracks were observed because of the REBCO tapes slitting process by many groups \cite{Cheggour2007,Yanagisawa2011,Chen2018,Gorospe2016,Isozaki2013,Mbaruku2008,Mitsui2010,Rogers2016,Rogers2017,Shin2019,Villaume2008}. Edge cracks due to the slitting process may cause the premature degradation when subjected to the tensile loading. Hence, the evaluation of edge cracks on the critical current ($I_{\rm c}$) performance under tensile stress is important for REBCO tapes application.

The influence of edge cracks on $I_{\rm c}$ performance under external stress was investigated by many groups. Cheggour \emph{et al} confirmed that the edge cracks caused by the slitting process remain stable and will not propagate when subjected to transverse compressive stress \cite{Cheggour2007}. However, the slitting edge has extremely low transverse tensile strength due to cracks and micro-peels under transverse tensile stress, and edge defects become the crack initiation \cite{Yanagisawa2011,2007van}. Liu \emph{et al} showed that REBCO tapes have higher shear strength under shear stress at the center and lower shear strength at the slitting edges \cite{2016Liu}. Isozaki \emph{et al} reported that the edge cracks grow and widen under fatigue tensile stresses \cite{Isozaki2013}, and draw the conclusion that influence of stress/strain on critical currents can be explained by assuming that some of edge cracks are widened and grown by the stress. Mbaruku \emph{et al} inferred that cracks initiated at the conductor edge and edge cracks served as crack nucleation sites during fatigue \cite{Mbaruku2008}. Mitsui \emph{et al} constructed a crack propagation model developed by the fatigue stresses \cite{Mitsui2010}, where the crack length will be increased by the repeated stresses. Rogers \emph{et al} concluded that the main sources of fatigue failure were the defects from the manufacturing process leading to crack propagation \cite{Rogers2016,Rogers2017}. Researchers mainly focused on edge cracks under fatigue tensile stress, and believed that egde cracks act as stress concentrators and grow during fatigue. However, few study about the edge cracks on $I_{\rm c}$ performance under quasi-static tensile stress was conducted. At present, researchers are in a state of speculation about the impact of edge cracks, and there is a lack of quantitative mechanical characterization methods based on fracture mechanics to assess the hazards of edge cracks.

The aim of this work is to evaluate the effect of edge cracks on the $I_{\rm c}$ performance under tensile stress. In section 3, $I_{\rm c}$ degradation under artificial cracks was conducted to evaluate the applicability of linear elastic fracture mechanics on the REBCO film. In section 4, a model was constructed to get the mixed stress intensity factor of multiple edge oblique cracks. The fracture toughness of the REBCO film was evaluated by nanoindentation. At last, multiple edge oblique cracks of two different REBCO tapes were evaluated.



\section{Experimental Details}


Four different tapes were fabricated by Shanghai Superconductor Technology Co., Ltd. (Shsctec) and Suzhou Advanced Materials Research Institution (Samri). As-received 10 mm-wide/12 mm-wide tapes from Shsctec/Samri were designated as Shsctec-10 mm/Samri-12 mm, where no edge cracks were observed. As-received 4 mm-wide/4 mm-wide tapes from Shsctec/Samri were designated as Shsctec-4 mm/Samri-4 mm, where obvious edge cracks were observed. The Shsctec-10 mm tapes were about 80 $\upmu$m total thickness and 10 mm wide. 3 $\upmu$m REBCO film was deposited on a template of 200 nm CeO$_2$ / 10 nm LaMnO$_3$ / 10 nm MgO / 25 nm Y$_2$O$_3$ / 80 nm Al$_2$O$_3$ / 50 $\upmu$m Hastelloy C-276. A 2 $\upmu$m Ag layer was used to enwrap the REBCO layer. Cu was sputtered onto the YBCO layer as a protective layer to a thickness of 10 $\upmu$m.

Shsctec-10 mm tapes were used to evaluate the applicability of linear elastic fracture mechanics for the REBCO film where there were no edge cracks. A femtosecond pulsed laser was used to fabricate artificial cracks on Shsctec-10 mm tapes and artificial cracks are perpendicular to load direction. Laser power and scanning rates are 25 mW and 50 $\upmu$m s$^{-1}$ with a defocusing distance -10 $\upmu$m. Then, tensile test and $I_{\rm c}$ measurement were conducted. The total length of the tensile specimens was 150 mm, and the initial gauge length of the sample was 120 mm. The bottom of the sample was fixed at the lower gripping holders. The other end was connected to the upper gripping holder which can move freely in the tensile strain direction. The in-suit $I_{\rm c}$ measurement of tapes under different tensile load were measured by using the four-point method in liquid nitrogen. 1 $\upmu$V/cm was defined as the criterion to determine $I_{\rm c}$, and the voltage sampling length was 2 cm.

The REBCO microstructures were observed under scanning electron microscopy (SEM), ZEISS GeminiSEM 300. Energy dispersive x-ray spectroscopy (EDS) was utilized to characterize the distribution of the elements. Nano indenter Keysight G200 was used to measure the fracture toughness of REBCO film.

\section{$I_{\rm c}$ degradation under artificial cracks}

For brittle materials with defects, linear elastic fracture mechanics is often used to analyze the critical conditions for the instability and cracks propagation. As a metal oxide ceramic, the REBCO film has the characteristic of brittleness. In order to validate the applicability of linear elastic fracture mechanics to REBCO film, we measured the critical degradation stress of REBCO tapes with artificial cracks of different scales. The correlation between artificial crack size and critical stress will be used to confirm the applicability of linear elastic fracture mechanics to REBCO materials.

\begin{figure}[H]
	\centering
	\includegraphics[width=0.99\textwidth]{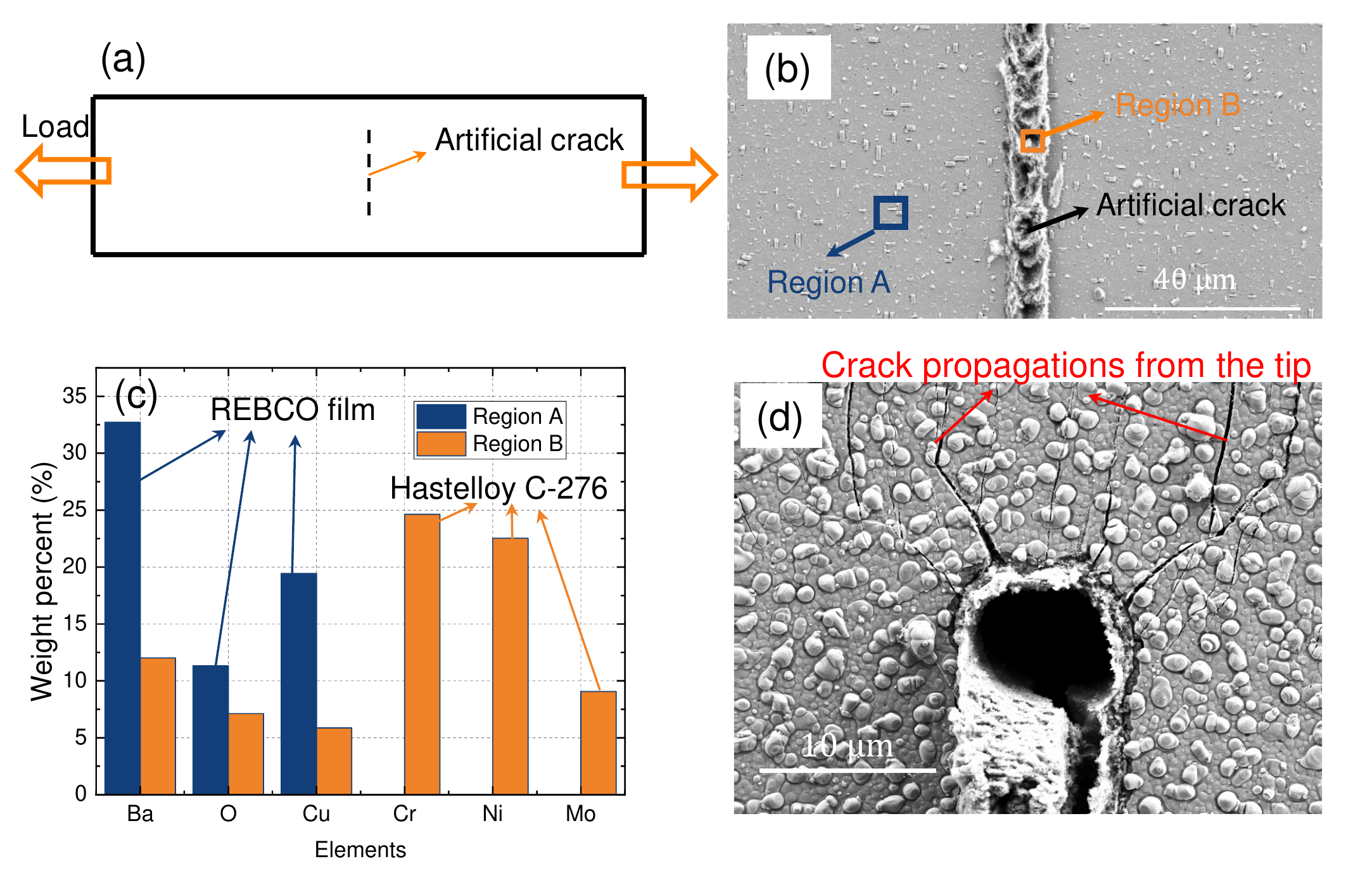}
	\caption{\textbf{(a) Illustration of artificial cracks fabricated by femtosecond pulsed laser. (b) Surface morphologies of artificial cracks. (c) EDS intensities of artificial cracks, suggesting that artificial cracks penetrate the REBCO film along thickness direction. (d) Crack propagations from the crack tip of the artificial crack, suggesting that artificial cracks has changed the stress field in fact.}}
	\label{Exp}
\end{figure}

Figure \ref{Exp}(a) is the illustration of artificial cracks fabricated by femtosecond pulsed laser. The artificial crack is prepared in the center of the tape, and the artificial crack is perpendicular to the load direction. Figure \ref{Exp}(b) shows the SEM images of the artificial crack fabricated on Shsctec-10 mm tapes. To accurately describe the crack depth, EDS intensities of region A and region B are shown in figure \ref{Exp}(c). Region A shows intensities for Ba, Cu ,and O, corresponding to the REBCO layer. Region B shows additional intensities of Cr, Ni ,and Mo, which are components of Hastelloy C-276. It suggests that the artificial crack penetrates the REBCO film along the thickness direction. As shown in figure \ref{Exp}(d), crack propagations from the crack tip of artificial crack are observed after the tensile test. It suggests that artificial cracks change the stress field in fact. Hence, artificial cracks can be used to evaluate the applicability of linear elastic fracture mechanics for REBCO film.

\begin{figure}[H]
	\centering
	\includegraphics[width=0.99\textwidth]{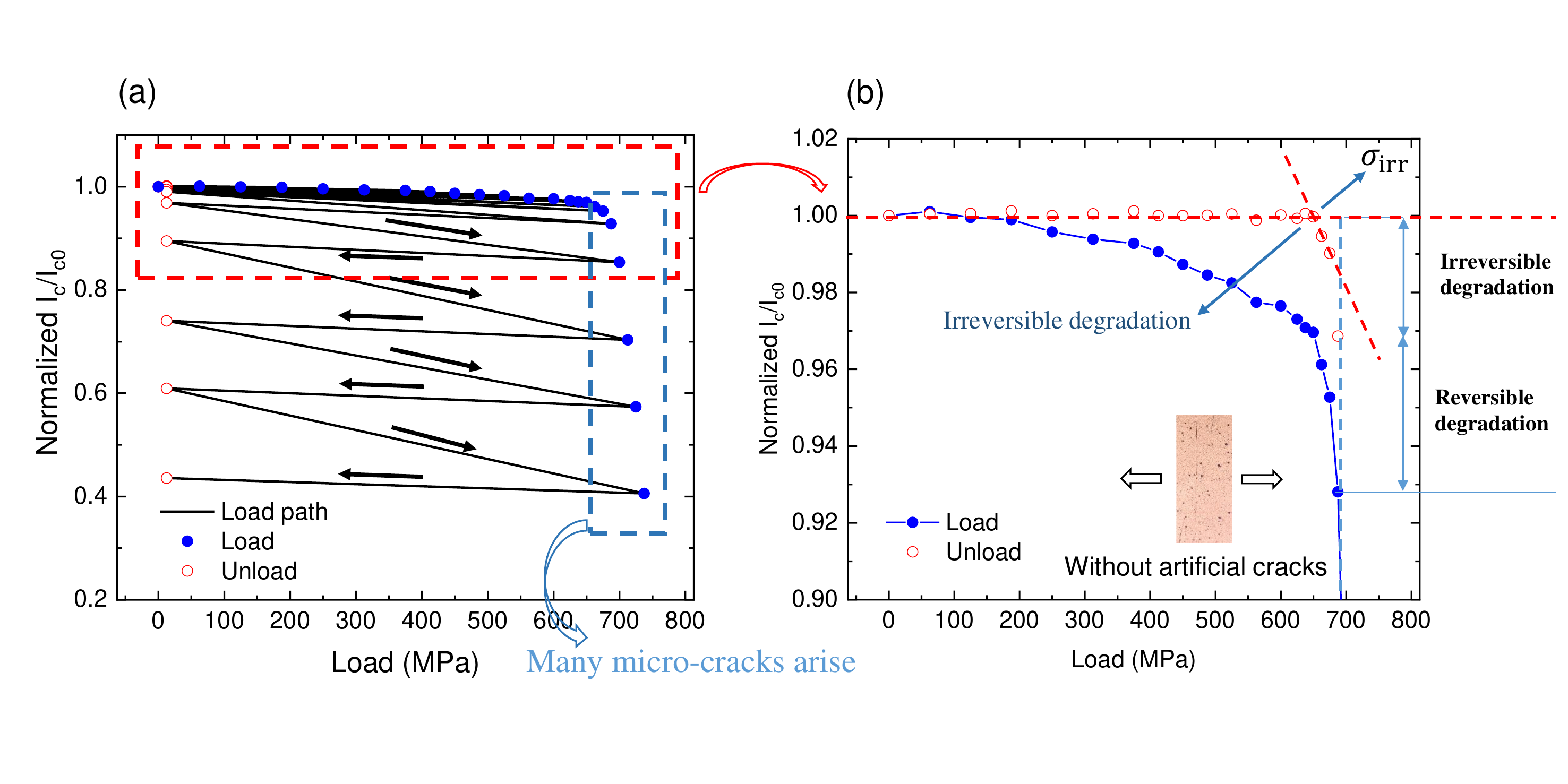}
	\caption{\textbf{$I_{\rm c}$ degradation of REBCO tapes without artificial cracks. (a) The load pathes of the $I_{\rm c}$ test. (b) The irreversible load $\sigma_{\rm irr}$, which is distinguished by the transition point of critical current in the unloaded state.}}
	\label{sds}
\end{figure}

We tested the critical properties $I_{\rm c}$ of REBCO tapes without artificial cracks under the loading and unloading conditions according to the load sequence shown in figure \ref{sds}(a). We are mainly concerned about the stage where the performance of REBCO tapes begins to show irreversible degradation as shown in the red box in figure \ref{sds}(a). Figure \ref{sds}(b) shows the performance degradation curve of REBCO tapes in this stage. It can be found that the performance degradation of REBCO tapes under load conditions is composed of two parts, one of which is reversible degradation (elastic deformation of REBCO lattice), and the other is irreversible degradation (generation of REBCO microcracks). In the unloaded conditions of REBCO tapes, the critical performance degradation is only determined by the generation of REBCO microcracks. Therefore, the performance transition point of REBCO tapes in the unloaded state is irreversible load $\sigma_{\rm irr}$. We can find that REBCO tapes without artificial defects appear irreversibly degraded at 663 MPa, which is caused by the intrinsic cracks in the REBCO film.

\begin{figure}[H]
	\centering
	\includegraphics[width=0.99\textwidth]{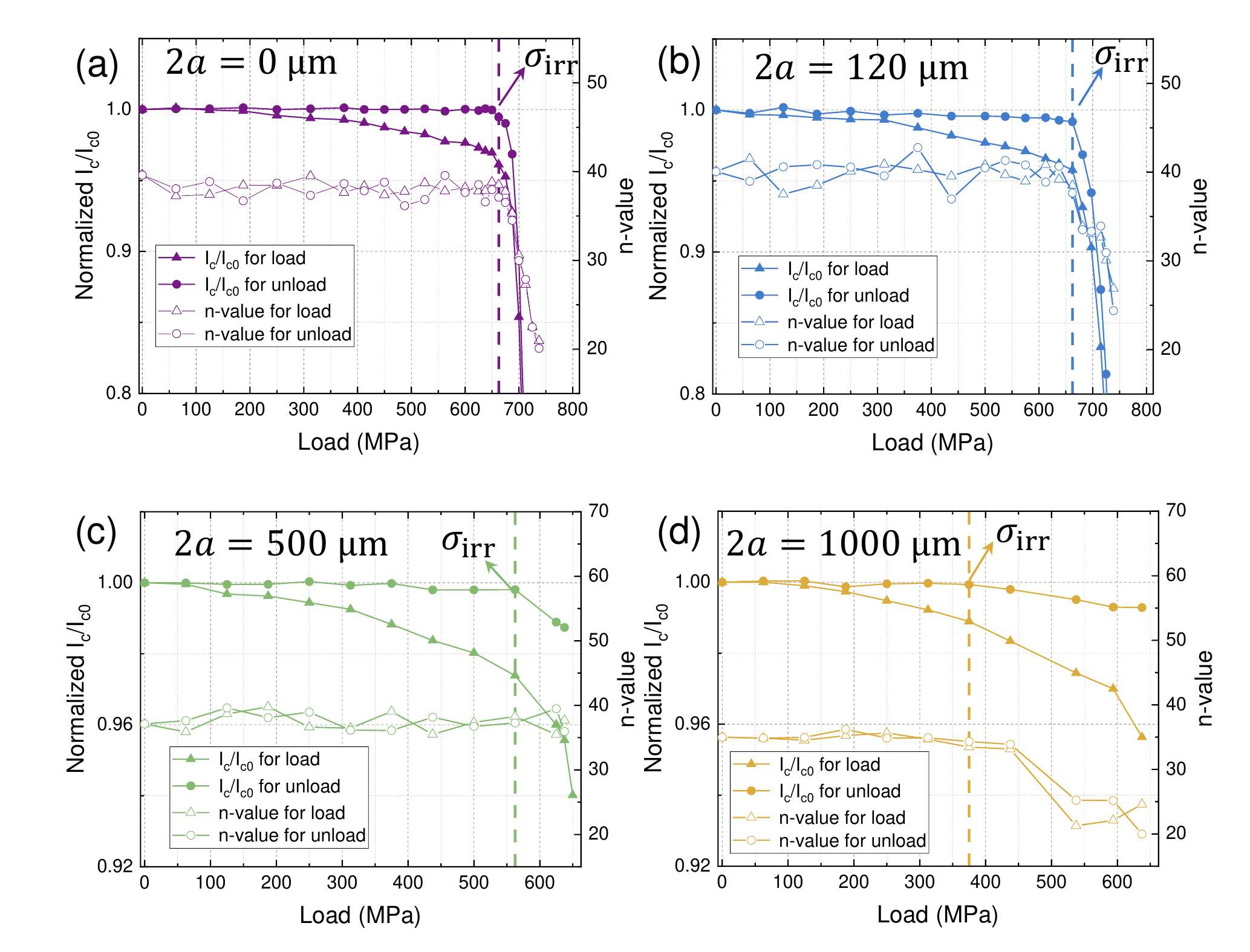}
	\caption{\textbf{$I_{\rm c}$ degradation under artificial cracks of (a) 0, (b) 120, (c) 500 and (d) 1000 $\upmu$m. The critical load defined by the $I_{\rm c}$ and $n-$value degradation were 663, 663, 563 and 375 MPa, respectively.}}
	\label{Icd}
\end{figure}

\begin{figure}[H]
	\centering
	\includegraphics[width=0.99\textwidth]{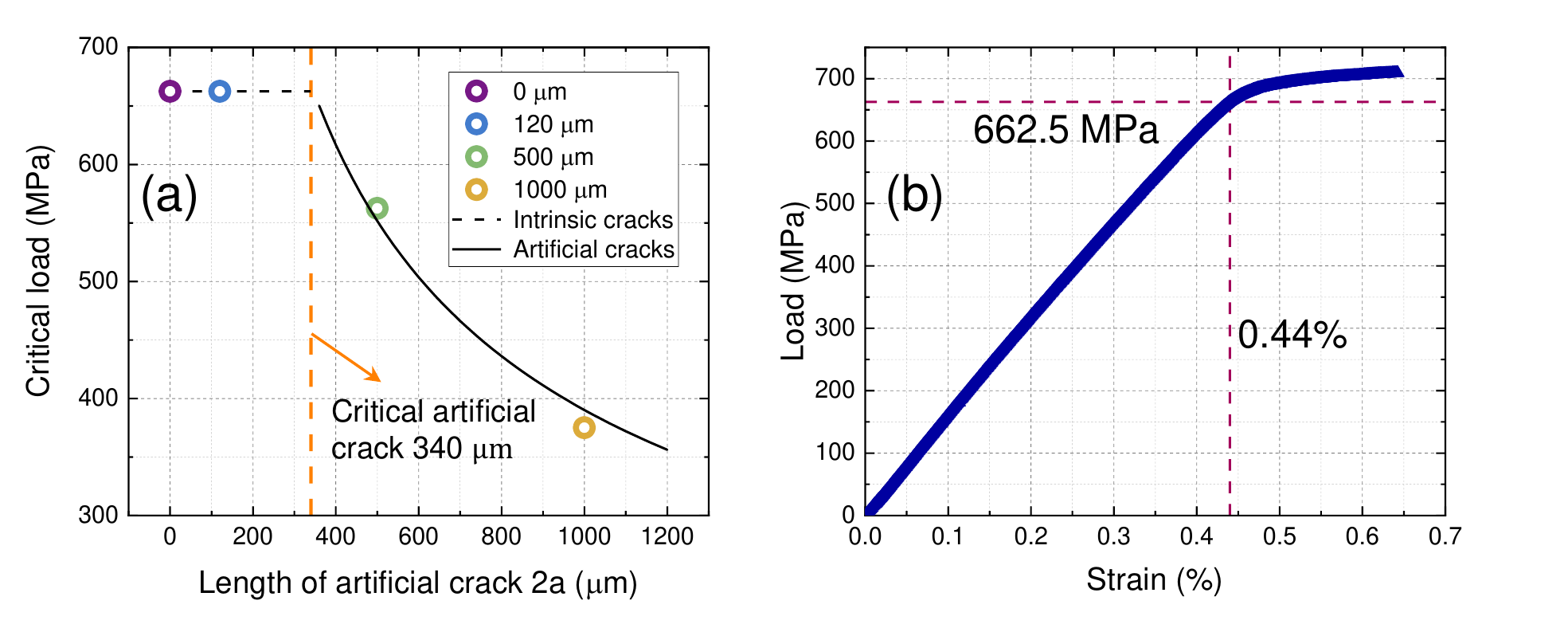}
	\caption{\textbf{(a) $\sigma_{\rm irr}-2a$ curve of artificial cracks. The $\sigma_{\rm irr}-2a$ curve can be divided into 2 stages. $I_{\rm c}$ degradation starts from internal intrinsic defects when the length of artificial cracks is less than 340 $\upmu$m; $I_{\rm c}$ degradation starts from artificial cracks when the length of artificial cracks is larger than 340 $\upmu$m.(b) Load-strain curve of tapes at 77K.}}
	\label{F2a}
\end{figure}

$I_{\rm c}$ degradations under tensile stress of artificial cracks 0, 120, 500 and 1000 $\upmu$m are shown in figure \ref{Icd}. $\sigma_{\rm irr}$ are 663, 663, 563 and 375 MPa for artificial cracks 0, 120, 500 and 1000 $\upmu$m, respectively. $\sigma_{\rm irr}-2a$ curves are shown in figure \ref{F2a}(a), and the circle dot is obtained from the $I_{\rm c}$ degradation curves in figure \ref{Icd}.

 The artificial crack is not an ideal crack. It is more like an inclusion, and there is a weak connection between the two crack surfaces. Hence, the length of artificial cracks $2a$ can't be inserted to standard stress intensity factor criterion $\sigma_{\rm irr}={K_{\rm IC}}/{ \sqrt{\pi a}}$.  Artificial cracks can be regarded as inclusions shown in figure \ref{Exp}(b), which can be described by De Kazinczy's model, where $\sigma_{\rm irr}={K_{\rm IC}}/{ \sqrt{k \pi a}}$  was used to evaluate the irreversible load of artificial cracks \cite{MURAKAMI1994163}. When the $2a$ of artificial cracks is less than 340 $\upmu$m, the $\sigma_{\rm irr}$ is 663 MPa, where no artificial cracks propagation occurs, and the degradation starts from the intrinsic cracks in REBCO film. Irreversible load  $\sigma_{\rm irr}$ is 663 MPa, which corresponds to irreversible strain 0.44\% shown in figure \ref{F2a}(b). When the $2a$ of artificial cracks is larger than 340 $\upmu$m, the $\sigma_{\rm irr}$ will decrease with the increase of $2a$, and the degradation is caused by crack propagation of artificial cracks. $\sigma_{\rm irr}$ is inversely proportional to $\sqrt{a}$, indicating that the linear elastic fracture mechanics can be applied for REBCO film.

\section{Results and discussions}

\subsection{The mixed stress intensity factor of multiple edge oblique cracks}

\begin{figure}[H]
	\centering
	\includegraphics[width=0.99\textwidth]{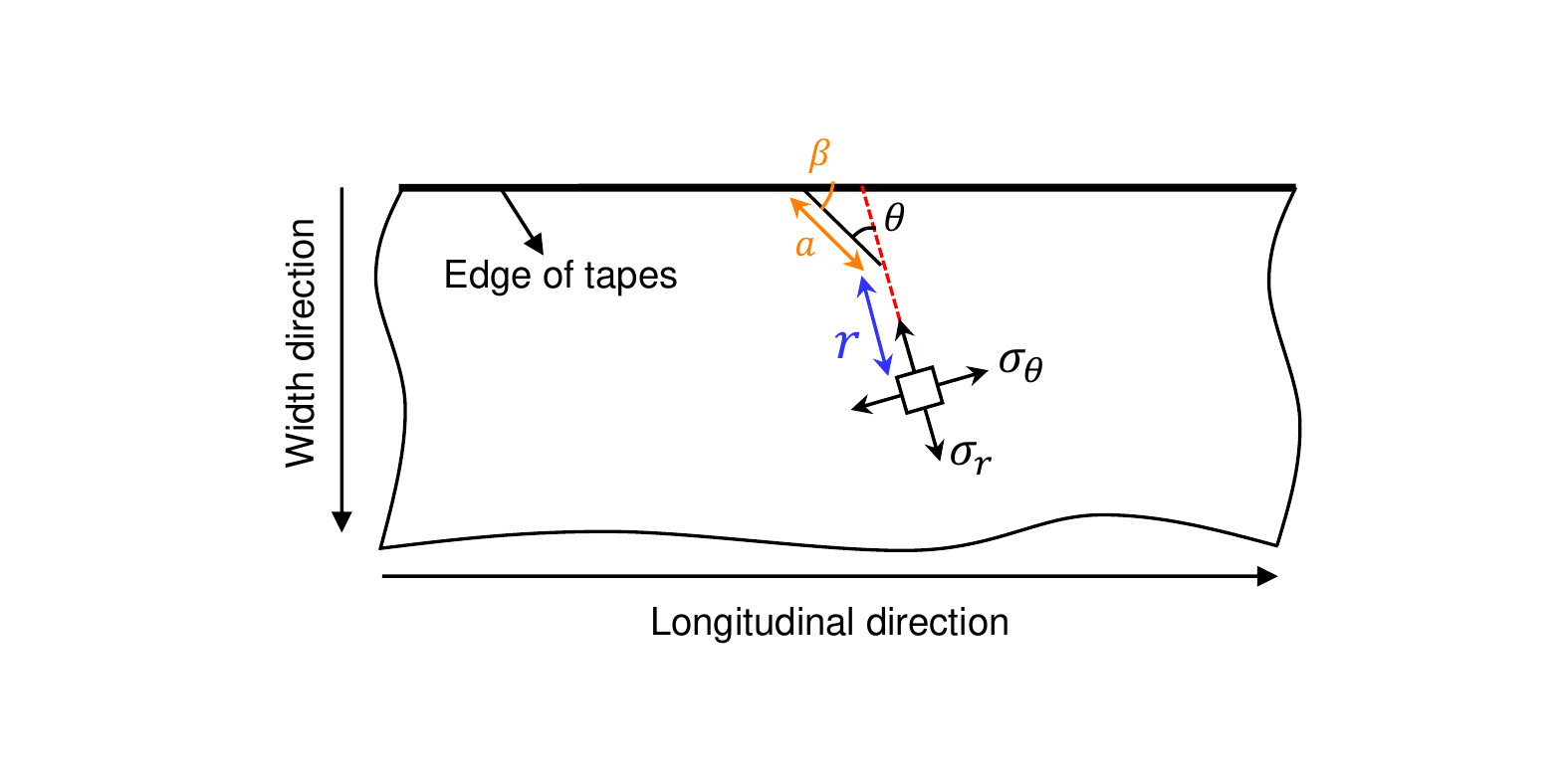}
	\caption{\textbf{Stress field distribution at the tip of the edge crack. The maximum normal stress criterion has been implemented in mixed-mode fracture mechanics as a fracture criterion.}}
	\label{shi}
\end{figure}

Oblique cracks in REBCO tapes are not only subjected to tensile stress but also shear stress, leading to a mixed-mode stress status shown in figure \ref{shi}. Hence, the polar stress components in an ideally elastic solid material under uniform and quasi-static loading are \cite{Erdogan1963}

\begin{equation}
	\begin{aligned}
		\sigma_{r}=\frac{1}{2 \sqrt{2 \pi r}}\left[K_{\rm I}(3+\cos \theta) \cos \frac{\theta}{2}+K_{\mathrm{II}}(3 \cos \theta-1) \sin \frac{\theta}{2}\right],\\
		\sigma_{\theta}=\frac{1}{2 \sqrt{2 \pi r}} \cos \frac{\theta}{2}\left[2 K_{\rm I} \cos ^{2} \frac{\theta}{2}-3 K_{\mathrm{II}} \sin \theta\right],\\
		\tau_{r \theta}=\frac{1}{2 \sqrt{2 \pi r}} \cos \frac{\theta}{2}\left[K_{\rm I} \sin \theta+3 K_{\mathrm{II}}(3 \cos \theta-1)\right],
	\end{aligned}
\end{equation} where $\sigma_r$ is the radial stress, $\sigma_\theta$ is the circumferential stress, $\tau_\theta$ is the shear stress, $K_{\rm I}$ is the stress intensity factor (SIF) for tensile crack, and $K_\mathrm{II}$ is the stress intensity factor for shear crack, respectively. The maximum normal stress criterion has been implemented in mixed-mode fracture mechanics as a fracture criterion because in many practical situations structural components experience tensile and shear loading simultaneously. The crack initiation angle $\theta_0$ and mixed stress intensity factor $K$ were derived by \cite{Erdogan1963}

\begin{equation}
	\begin{aligned}
		\left(\frac{\partial \sigma_{\theta}}{\partial \theta}\right)=0,\\
		\theta_{0}=-\arccos \frac{3 K_{\mathrm{II}}^{2}+\sqrt{K_{\rm I}^{4}+8 K_{\rm I}^{2} K_{\mathrm{II}}^{2}}}{K_{\mathrm{I}}^{2}+9 K_{\mathrm{II}}^{2}},\\
		K=\cos \frac{\theta_{0}}{2}\left[K_{\rm I} \cos ^{2} \frac{\theta_{0}}{2}-\frac{3}{2} K_\mathrm{II} \sin \theta_{0}\right],
	\end{aligned}
\end{equation}where $K$ is the mixed-mode stress intensity factor of cracks, and $\theta_0$ is the fracture angle which is perpendicular to the maximum principal tensile stress. With the principal stress theory, the ${K_{\rm I}}$ and ${K_{\rm II}}$ of edge cracks should be obtained to solve $K$ and $\theta_0$.

\begin{figure}[H]
	\centering
	\includegraphics[width=0.99\textwidth]{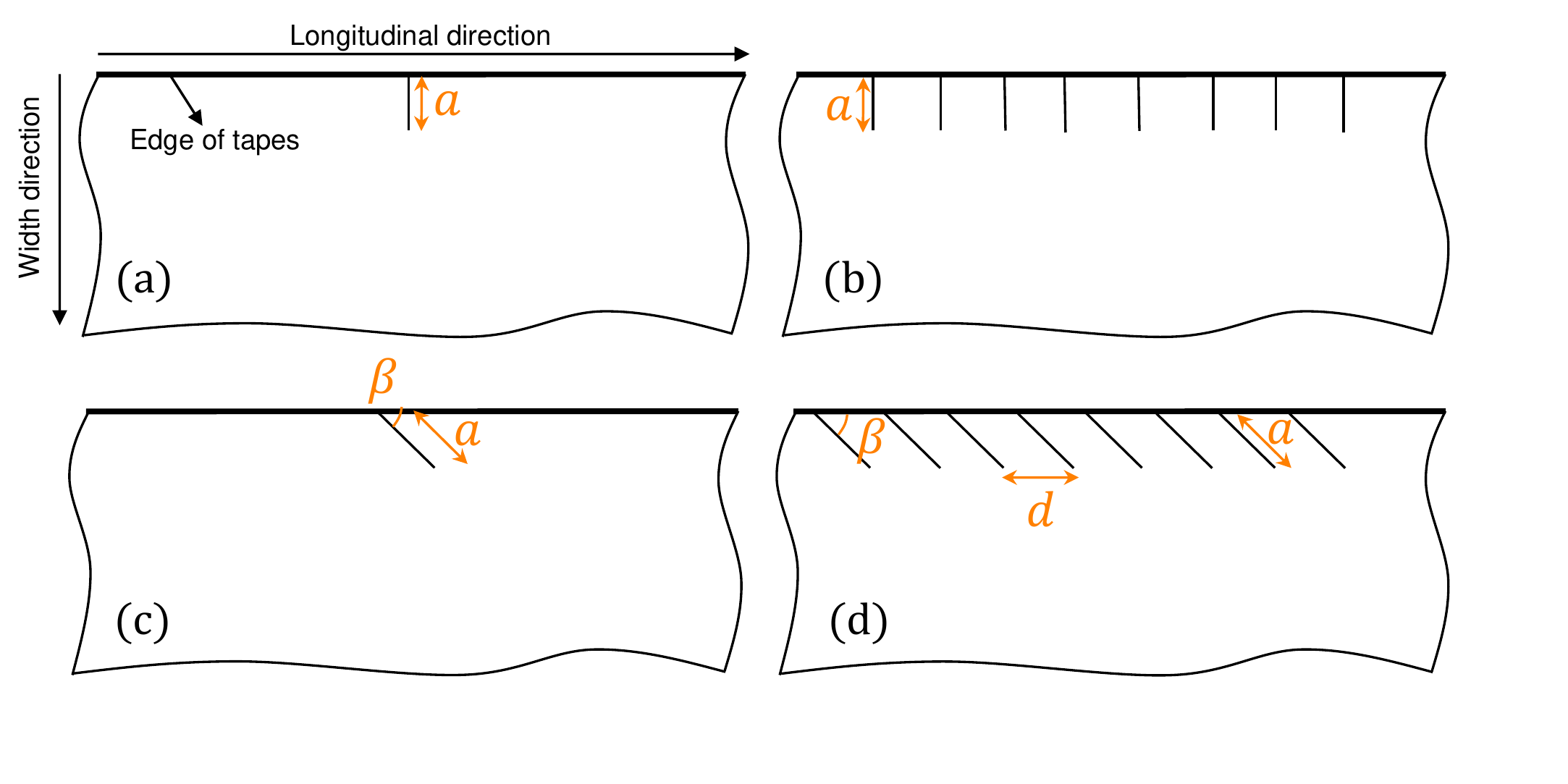}
	\caption{\textbf{Four different edge cracks. (a) Single edge perpendicular crack; (b) multiple edge perpendicular cracks; (c) single edge oblique crack; (d) multiple edge oblique cracks. The ${K_{\rm I}}$ and ${K_{\rm II}}$ for (a)-(c) can be obtained in reference, and a mapping method was used to calculate the ${K_{\rm I}}$ and ${K_{\rm II}}$ for (d) in this work.}}
	\label{K}
\end{figure}

The stress intensity factor of single edge perpendicular crack and multiple edge perpendicular cracks shown in figure \ref{K}(a) and figure \ref{K}(b) are from reference \cite{Tada2000,Sih1973}, where  the ${K_{\rm I}}$ and ${K_{\rm II}}$ were calculated as

\begin{equation}
\begin{array}{l}
	K_{\mathrm{I,a}}=F_{1} \sigma \sqrt{\pi a}, K_{\mathrm{II,a}}=0\\
\end{array}
\end{equation}and

\begin{equation}
\begin{array}{l}
K_{\mathrm{I,b}}=F_{2} \sigma \sqrt{\pi a}, K_{\mathrm{II,b}}=0,\\
\end{array}
\end{equation}where $\mathrm{F_1}=1.1215$ \cite{Tada2000}. Considering multiple edge cracks of depth $a$ and spacing $d$, the $\mathrm{F_{2}}$ is

\begin{equation}
\begin{array}{r}

F_2=\frac{1}{\sqrt{\pi}}\sqrt{\frac{b}{s}}\left[1+\frac{1}{2} \frac{b}{s}+\frac{3}{8} \frac{b^{2}}{s^{2}}+\frac{5}{16} \frac{b^{3}}{s^{3}}+\frac{35}{128} \frac{b^{4}}{s^{4}}+\frac{63}{256} \frac{b^{5}}{s^{5}}+\frac{231}{1024} \frac{b^{6}}{s^{6}}\right]+ \\
\sqrt{\frac{b}{s}}\left[22.501 \frac{b^{7}}{s^{7}}-63.502 \frac{b^{8}}{s^{8}}+58.045 \frac{b^{9}}{s^{9}}-17.577 \frac{b^{10}}{s^{10}}\right],\\
\end{array}
\end{equation} where $b=d/2$ and $s=a+d/2$ \cite{Sih1973}.

\begin{figure}[H]
	\centering
	\includegraphics[width=0.99\textwidth]{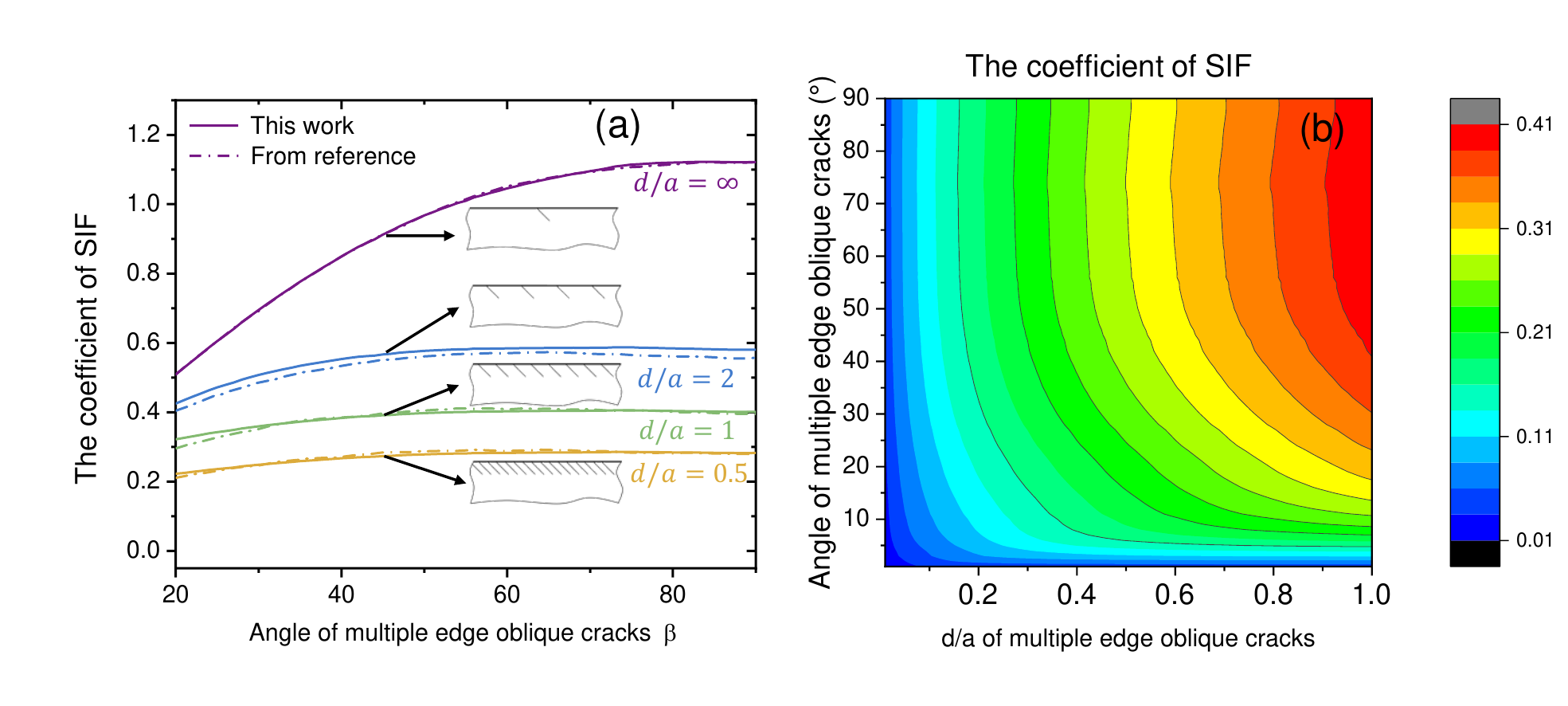}
	\caption{\textbf{(a) The coefficient of stress intensity factors of multiple edge oblique cracks at $d/a=\infty, 2, 1$ and $0.5$ with different angles compared with reference \cite{Jin2006}. (b) The coefficient of stress intensity factors of multiple edge oblique cracks with different $\beta$ and $d/a$.}}
	\label{SIF}
\end{figure}

There is also a numerical solution of a single edge oblique crack shown in figure \ref{K}(c) \cite{Hasebe1980}, where  the ${K_{\rm I}}$ and ${K_{\rm II}}$ were calculated as

\begin{equation}
\begin{array}{l}
	K_{\mathrm{I,c}}=F_{3} \sigma \sqrt{\pi a}, K_{\mathrm{II,c}}=F_{4} \sigma \sqrt{\pi a}.\\
\end{array}
\end{equation}where both $F_{3}$ and $F_{4}$ are related to the angle $\beta$. However, there is no enough results of stress intensity factor in multiple edge oblique cracks shown in figure \ref{K}(d)\cite{Jin2006}. For a row of multiple edge perpendicular cracks of a semi-infinite plate, the stress intensity factor can be characterized by $F_{2}$. When the cracks are sparse, the result is consistent with the stress intensity factor $F_{1}$ of a single edge perpendicular crack. Therefore, it can be considered that there is a conversion coefficient between the stress intensity factor of multiple edge perpendicular cracks and the stress intensity factor of a single edge perpendicular crack. The coefficient is $F_{2}/F_{1}$, which is related to the crack properties. For a row of oblique edge cracks, considering the conversion factor between the single crack and the crack array introduced by the crack properties, the type I and type II stress intensity factors of the oblique edge crack array can be obtained respectively as

\begin{equation}
{{K_{\rm I,d}}}=F_{3} \sigma \sqrt{\pi a} F_{2} / F_{1},
\end{equation}
 
\begin{equation}
{{K_{\rm II,d}}}=F_{4} \sigma \sqrt{\pi a} F_{2} / F_{1}.
\end{equation}
For ease of description, the mixed stress intensity factor of multiple edge oblique cracks is

\begin{equation}
	{{K_{\rm d}}}=\cos \frac{\theta_{0}}{2}\left[K_\mathrm{I,d} \cos ^{2} \frac{\theta_{0}}{2}-\frac{3}{2} K_\mathrm{II,d} \sin \theta_{0}\right] \rightarrow 	{{K_{\rm d}}}=F\sigma \sqrt{\pi a},
\end{equation} where $F$ is the coefficient of stress intensity factor. Hence, the stress intensity factor of multiple edge oblique cracks with different angles, lengths, and spacings can be derived by equation(9). $F$ of edge cracks at $d/a=\infty, 2, 1$, and $0.5$ with different angles is shown in figure \ref{SIF}(a), which is in reasonable agreement with that in the references \cite{Jin2006}. The stress intensity factors of multiple edge oblique cracks are shown in figure \ref{SIF}(b), where the angle of cracks is $\beta$, the length of cracks is $a$, and the spacing of cracks is $d$. So, if we get the fracture toughness $K_{\rm IC}$ and the properties of edge cracks, such as $\beta$, $a$, and $d$, the evaluation of the multiple edge oblique cracks on mechanical properties can be conducted.

\subsection{The fracture toughness evaluation based on nanoindentation}


\begin{figure}[H]
	\centering
	\includegraphics[width=0.99\textwidth]{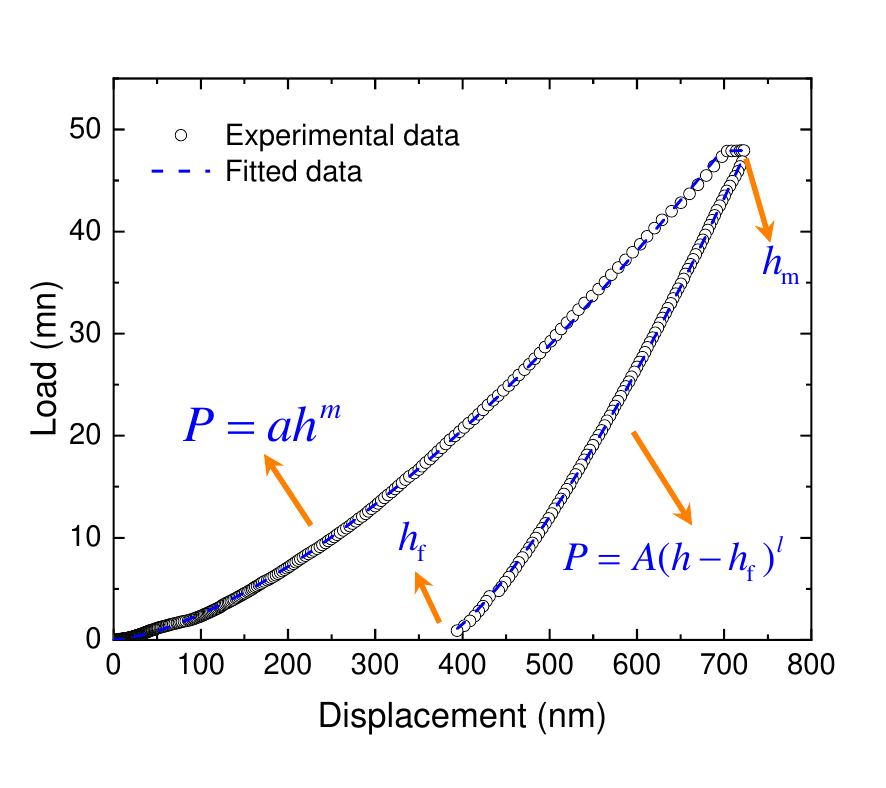}
	\caption{\textbf{The load-displacement curve of the REBCO film using a Berkovich indenter. The fracture toughness of the REBCO film could be evaluated.}}
	\label{Nano}
\end{figure}

The nanoindentation can be used to evaluate the fracture toughness of materials \cite{Cheng2002}. Figure \ref{Nano} shows the load-displacement curve of REBCO film using a Berkovich indenter, the respective loading and unloading curves can be given by $P=ah^m$ and $P=b(h-h_{\rm f})^l$, where  $h_{\rm f}$ is the residual final depth, and $h_{\rm m}$ is the maximum indentation depth. The fracture work can be calculated by

\begin{equation}
	W_{\rm c}=W_{\rm T}-W_{\rm p}-W_{\rm e},
\end{equation} where $W_\mathrm T$ is the total work, $W_\mathrm p$ is the plastic work, $W_\mathrm e$ is the elastic work, and $W_\mathrm c$ is the fracture work. The total work $W_\mathrm T$ and the elastic work $W_\mathrm e$ can be obtained by the load-displacement curve. The plastic work $W_\mathrm p$ is then given by \cite{Mencik1994}

\begin{equation}
	\frac{W_{\rm p}}{W_{\mathrm{T}}}=\frac{m+1}{l+1} \frac{h_{\mathrm{f}}}{h_{\mathrm{m}}}-\frac{m-l}{l+1}.
\end{equation}

Therefore the critical energy release rate $G_{\rm c}$ can then be determined as $G_{\rm c}={W_{\rm c}}/{A_{\rm m}}$, where $A_{\rm m}$ is the maximum crack area, calculated as $A_{\max }=24.5 h_{m}^{2}$ for Berkovich indenter. The fracture toughness of REBCO film can be obtained $K_{\rm IC}=\sqrt{G_{\rm c}E_{\rm r}}$, where $E_{\rm r}$ is the equivalent modulus of nanoindentation. $K_{\rm IC}=2.3 \pm 0.6 $ MPa m$^{-1/2}$ of REBCO film is obtained after 20 nanoindentation tests.

\subsection{Evaluation of multiple edge oblique cracks}

\begin{figure}[H]
	\centering
	\includegraphics[width=0.99\textwidth]{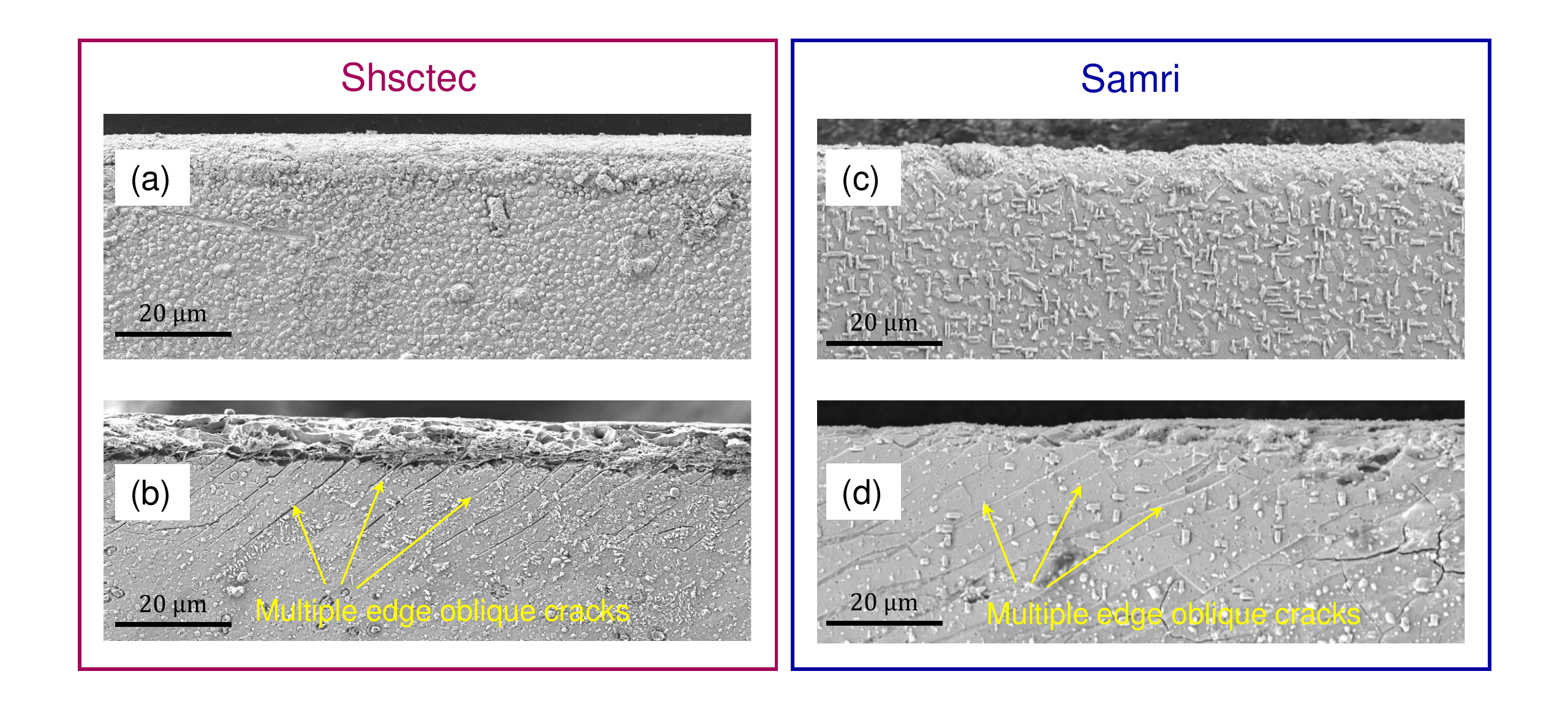}
	\caption{\textbf{Surface morphologies of REBCO films. (a) The edge of Shsctec-10 mm; (b) the slitting edge of Shsctec-4 mm; (c) the edge of Smari-12 mm. (d) the slitting edge of Samri-4 mm. Edge cracks were not observed in no-slitting tapes and multiple edge oblique cracks were induced in the slitting process of HTS tapes.}}
	\label{sample}
\end{figure}

\begin{figure}[H]
	\centering
	\includegraphics[width=0.99\textwidth]{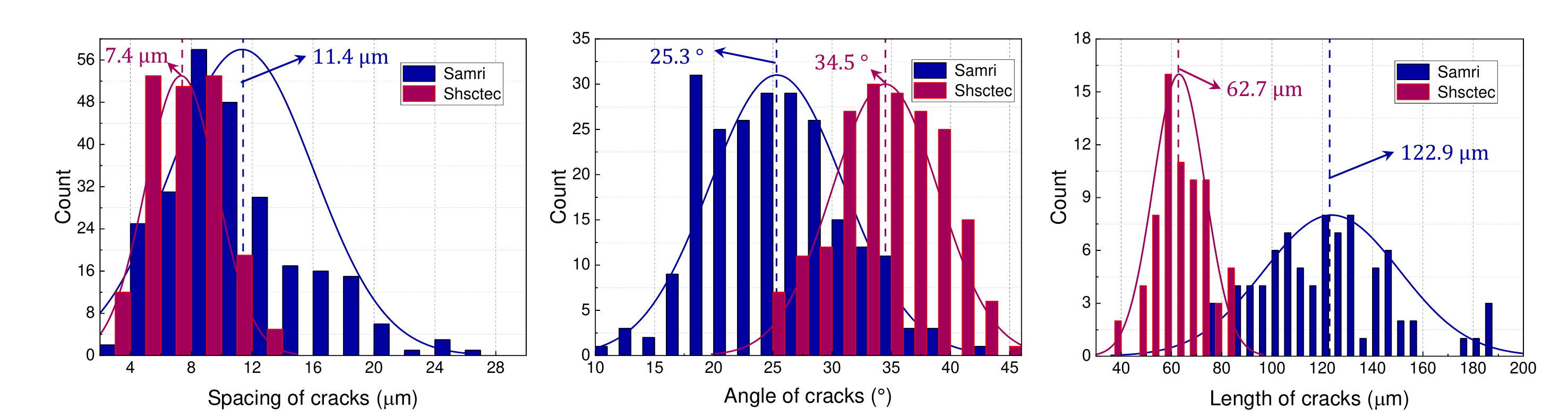}
	\caption{\textbf{Crack properties spacing $d$, angle $\beta$ and length $a$ of tapes fabricated by Samri and Shsctec. Crack characteristics follow the normal distributions.}}
	\label{DATA}
\end{figure}

SEM images of four different REBCO tapes are shown in figure \ref{sample}. There is no edge cracks observed in no-slitting HTS tapes Shsctec-10 mm and Samri-12 mm. However, multiple edge oblique cracks are induced in the slitting process of Shsctec-4 mm and Samri-4 mm. The angle of cracks $\beta$, the length of cracks $a$, and the spacing of cracks $d$ are displayed in figure \ref{DATA}. From the frequency distribution of crack properties, the dispersed data can be described effectively by a normal distribution. The crack properties of tapes fabricated by Samri are $d=11.4 \pm 0.4$ $\mu \mathrm{m}$, $\beta=25.3 \pm 0.3$ $^{\circ}$ and $a=122.9 \pm 2.2$ $\mu \mathrm{m}$. The crack properties of tapes fabricated by Shsctec are $d=7.4 \pm 0.1$ $\mu \mathrm{m}$, $\beta=34.5 \pm 0.5$ $^{\circ}$ and $a=62.7 \pm 1.5$ $\mu \mathrm{m}$.

For the tapes without artificial cracks, the irreversible load is 663 MPa, corresponding to 0.44\% strain. Actually, 663 MPa is the average stress inside the REBCO tapes, and the stress in the REBCO layer is often higher. Therefore, we calculate the product of strain and modulus to find that the stress inside the REBCO, which is 700MPa at 0.44\% strain. The elastic modulus of REBCO film is 159 GPa. For REBCO tapes without edge cracks, when the applied load is less than 700 MPa, the REBCO tapes will not suffer from intrinsic damage; when the applied load is greater than 700 MPa, the REBCO tapes will suffer from intrinsic damage. Therefore, we check the edge cracks of REBCO tapes under a load of 700 MPa. When the edge cracks do not propagate at 700 MPa, we believe that the edge crack does not affect the critical load of the REBCO tapes and will not degrade the critical current before 0.44\% strain. When the REBCO edge crack expands at 700 MPa, it is considered that the edge crack will reduce the critical stress of the REBCO tapes and cause premature degradation. We judge the influence of each type of crack based on this idea.

\begin{figure}[H]
	\centering
	\includegraphics[width=0.99\textwidth]{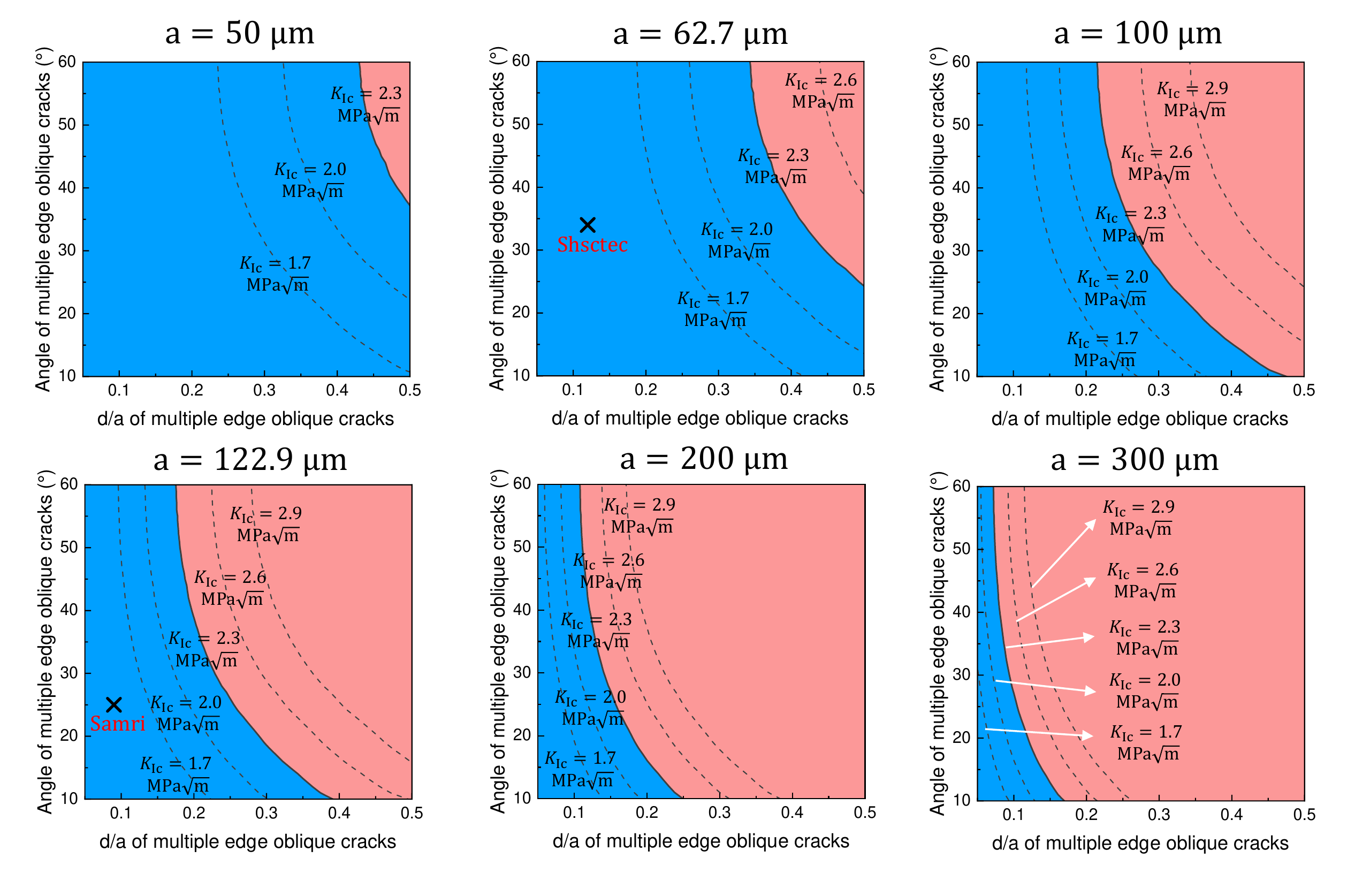}
	\caption{\textbf{Fracture limit diagram for multiple edge oblique cracks. The edge cracks in the blue zone are in a stable state and will not propagate. The edge cracks in the red area are in a dangerous state.}}
	\label{Sta}
\end{figure}

We evaluated the fracture limit diagrams of edge cracks of different lengths, angles, and spacings, as shown in figure \ref{Sta}. For example, when the edge cracks length is 50 $\mu \mathrm{m}$, its different crack angle and spacing divide the diagram into two regions. The edge cracks in the blue zone are in a stable state and will not propagate. The edge cracks in the red area are in a dangerous state, which will expand and the $I_{\rm c}$ will degrade before 0.44\% strain. It can be found that the smaller the crack length, the more secure the edge cracks. The sparser and larger the angle of the crack array will make the edge cracks tend to be dangerous. The fracture toughness of REBCO film is 2.3 MPa $\sqrt{\mathrm{m}}$. When we increase the fracture toughness of the REBCO, it can be found that the stable region will be expanded. The crack properties of tapes fabricated by Samri and Shsctec are located in the safe region. Hence, multiple edge oblique cracks in both tapes will not degrade the $I_{\rm c}$ performance before the irreversible strain 0.44\% strain in this work.


Angle $\beta$, length $a$, and spacing $d$ play important roles in the stress intensity factor. Edge cracks with a smaller angle, a shorter length, and a higher density can tolerate a higher stress field. Hence, the properties of edge cracks should be characterized, which is helpful to evaluate and improve the manufacturing process of REBCO tapes. In this work, edge cracks of two tapes show no influence on the $I_{\rm c}$ performance before the irreversible strain 0.44\% strain. Nevertheless, the results obtained do not signify that edge cracks are harmless to the conductor performance under any circumstances. 


\section{Conclusions}

$I_{\rm c}$ degradation under artificial cracks was prepared to confirm the applicability of linear elastic fracture mechanics for REBCO film. $\sigma_{\rm irr}$ is 663, 663, 563, and 375 MPa for artificial cracks 0, 120, 500, and 1000 $\upmu$m, respectively. It indicated that the linear elastic fracture mechanics can be applied for REBCO film. A model considering edge crack properties angle $\beta$, spacing $d$ and length $a$ is constructed to evaluate the stress intensity factor with good predictability. $K_{\rm IC}=2.3 \pm 0.6 $ MPa m$^{-1/2}$ of REBCO film in this work is obtained. The crack properties of tapes are $d=11.4 \pm 0.4$ $\mu \mathrm{m}$, $\beta=25.3 \pm 0.3$ $^{\circ}$ and $a=122.9 \pm 2.2$ $\mu \mathrm{m}$ for Samri and $d=7.4 \pm 0.1$ $\mu \mathrm{m}$, $\beta=34.5 \pm 0.5$ $^{\circ}$ and $a=62.7 \pm 1.5$ $\mu \mathrm{m}$ for Shsctec. The crack properties of tapes fabricated by Samri and Shsctec are located in the safe region, indicating that edge cracks will not degrade the $I_{\rm c}$ performance before the irreversible strain 0.44\%. This model could be applied to evaluate the operation range and improve the manufacturing process of REBCO tapes.

\section{Acknowledgments}

This work was supported by the Strategic Priority Research Program of Chinese
Academy of Sciences under Grant No. XDB25000000, National Natural Science
Foundation of China under Grant No. 52007089 and the Tribology Science Fund of
State Key Laboratory of Tribology, China under Grant No. SKLT2020B01.

\section{References}

\bibliography{iop}

\end{document}